\documentclass[showpacs,preprintnumbers,twocolumn,amsmath,amssymb, groupedaddress,superscriptaddress]{revtex4-1}
\usepackage{graphicx}
\usepackage{dcolumn}
\usepackage{bm}

\begin{document}

\title{Microscopic analysis of $^{10,11}$Be elastic scattering on protons and
nuclei and breakup processes of $^{11}$Be within the $^{10}$Be+$n$
cluster model}

\author{V.~K.~Lukyanov}
\affiliation{Joint Institute for Nuclear Research, Dubna 141980,
Russia}

\author{D.~N.~Kadrev}
\affiliation{Institute for Nuclear Research and Nuclear Energy,
Bulgarian Academy of Sciences, Sofia 1784, Bulgaria}

\author{E.~V.~Zemlyanaya}
\affiliation{Joint Institute for Nuclear Research, Dubna 141980,
Russia}

\author{K.~Spasova}
\affiliation{Institute for Nuclear Research and Nuclear Energy,
Bulgarian Academy of Sciences, Sofia 1784, Bulgaria}
\affiliation{University "Ep.~K. Preslavski", Shumen 9712,
Bulgaria}

\author{K.~V.~Lukyanov}
\affiliation{Joint Institute for Nuclear Research, Dubna 141980,
Russia}

\author{A.~N.~Antonov}
\affiliation{Institute for Nuclear Research and Nuclear Energy,
Bulgarian Academy of Sciences, Sofia 1784, Bulgaria}

\author{M.~K.~Gaidarov}
\affiliation{Institute for Nuclear Research and Nuclear Energy,
Bulgarian Academy of Sciences, Sofia 1784, Bulgaria}

\begin{abstract}
The density distributions of $^{10}$Be and $^{11}$Be nuclei
obtained within the quantum Monte Carlo (QMC) model and the
generator coordinate method (GCM) are used to calculate the
microscopic optical potentials (OPs) and cross sections of elastic
scattering of these nuclei on protons and $^{12}$C at energies
$E<100$ MeV/nucleon. The real part of the OP is calculated using
the folding model with the exchange terms included, while the
imaginary part of the OP that reproduces the phase of scattering
is obtained in the high-energy approximation (HEA). In this hybrid
model of OP the free parameters are the depths of the real and
imaginary parts obtained by fitting the experimental data. The
well known energy dependence of the volume integrals is used as a
physical constraint to resolve the ambiguities of the parameter
values. The role of the spin-orbit potential and the surface
contribution to the OP is studied for an adequate description of
available experimental elastic scattering cross section data.
Also, the cluster model, in which $^{11}$Be consists of a $n$-halo
and the $^{10}$Be core, is adopted. Within the latter, the breakup
cross sections of $^{11}$Be nucleus on $^{9}$Be, $^{93}$Nb,
$^{181}$Ta, and $^{238}$U targets and momentum distributions of
$^{10}$Be fragments are calculated and compared with the existing
experimental data.
\end{abstract}


\maketitle

\section{Introduction
\label{s:intro}}

The discovery of halo nuclei \cite{Tanihata85a} has been related
to the measured interaction cross sections of nuclei like
$^{6,8}$He, $^{11}$Li, Be isotopes with various target nuclei
\cite{Tanihata85b,Tanihata85b,Tanihata88,Mittig87,Tanihata95,
Hansen95}. The evidences of the existence of an extended halo in
neutron-rich nuclei are based on the observed unusually narrow
momentum distribution of a core fragment and enhanced reaction
cross section. The first example was the breakup of $^{11}$Li at
high energies \cite{Kobayashi88,Bertsch89,Anne90,Esbensen91} by
observing the large interaction reaction cross section
\cite{Tanihata85b} and the narrow momentum distribution of
$^{9}$Li in the breakup of $^{11}$Li, e.g., in the reaction
$^{11}$Li+$^{12}$C at $E=800$ MeV/nucleon \cite{Kobayashi88}. Here
we should mention also the results of the experiments at lower
energies ($E=60$ MeV/nucleon) of scattering of $^{11}$Li on
$^{9}$Be, $^{93}$Nb and $^{181}$Ta \cite{Orr92} and of $^{11}$Li
on a wide range of nuclei from $^{9}$Be to $^{238}$U \cite{Orr95}.
As shown in Ref.~\cite{Kelley95}, not only scattering but also the
breakup of $^{11}$Be in the collisions with the target nuclei
$^{93}$Nb, $^{181}$Ta, and $^{238}$U play a decisive role when
studying the internal cluster structure of $^{11}$Be. Indeed, the
narrow peak of the momentum distributions of the breakup fragments
of such a neutron-rich nucleus reflects the very large extension
of its wave function, compared to that of the core nucleus
$^{10}$Be, and thus evidences the existence of the nuclear halo
\cite{Baye2012,Barranco96,Hencken96,Bertulani2004,Bertulani92,Ershov2004,Bertulani2002}.
As was concluded in \cite{Bertulani92}, namely the longitudinal
component of the momentum (taken along the beam or the
$z$-direction) provides the most accurate information on the
intrinsic properties of the halo, being insensitive to details of
the collision and the size of the target. In addition, recent
measurements of the charge radii of $^{7,9,10,11}$Be pointed out
that the average distance between the halo neutrons and the
$^{10}$Be dense core of the $^{11}$Be nucleus is around 7 fm
\cite{Nortershauser2009}. Thus, the halo neutron is about three
times as far from the dense core as is the outermost proton
because the core itself has a radius of only 2.5 fm.

An important finding when investigating reactions with $^{10}$Be
and $^{11}$Be nuclei, in particular the $^{10}$Be+$n$ breakup of
$^{11}$Be, is the effect of the deformed $^{10}$Be core on the
two-body cluster structure of $^{11}$Be. In fact, in the $^{11}$Be
nucleus the inversion of the $p_{1/2}$ and $s_{1/2}$ orbitals
predicted by Talmi and Unna \cite{Talmi60} and confirmed by
Alburger {\it et al.} \cite{Alburger64} leads to a 1/2$^{+}$
ground state. Also, the probability of the E1 transition from this
ground state to the 1/2$^{-}$ first excited state of $^{11}$Be
located at 320 keV excitation energy is the largest ever measured
in light nuclei \cite{Millener83,Kelley2012}. The effects of the
core deformation on the breakup of $^{11}$Be on protons have been
studied in several works. For example, in addition to a two-body
cluster structure with an inert $^{10}$Be(0$^{+}$) core and a
valence neutron used in Ref.~\cite{Shrivastava2004} in the
continuum discretised coupled-channels (CDCC) calculations of
elastic and inelastic proton scattering on $^{11}$Be, the authors
have also discussed the necessity to account for contributions
from configurations involving excited states of the $^{10}$Be core
to the $^{10}$Be+$n$ continuum of $^{11}$Be. Crespo {\it et al.}
\cite{Crespo2011} have found that the core excitation
$p$+$^{10}$Be(0$_{1}^{+}$) $\rightarrow $
$p$+$^{10}$Be(2$_{1}^{+}$) provides a significant contribution to
the breakup cross section of $^{11}$Be on the proton target at
63.7 MeV/nucleon incident energy.

In the earlier works (e.g., Ref.~\cite{Cortina-Gil97}) the elastic
scattering cross sections of $^{10,11}$Be on protons have been
calculated using phenomenological OPs of given forms with numerous
fitting parameters of their real (ReOP) and imaginary (ImOP)
parts. However, in the further calculations the more physically
motivated microscopic folding models were applied (see, e.g.,
\cite{Amos2005,Satchler79,Khoa1993,Khoa2000}). In many works
\cite{Satchler79,Khoa1993,Khoa2000} the folding procedure was
explored for the real part of the OP. Within the latter procedure
the direct and exchange parts of the ReOP with effective
nucleon-nucleon forces are calculated. At the same time the P is
usually taken in a phenomenological form. Many successful
applications of this model have been made for the proton- and
nucleus-nucleus collisions (see, e.g., cycles of works
\cite{Khoa1993,Khoa97,Avrigeanu2000}). This model was also
explored in Refs.~\cite{Hassan2009,Farag2012} for scattering of
$^{10,11}$Be+$p$, where the exchange part of the folded ReOP is
taken in the form of the zero-range prescription and the density
distribution of $^{11}$Be has the Gaussian-oscillator form. In the
recent work \cite{Farag2014a} the authors account for the full
exchange part of the ReOP, while the ImOP was calculated using the
folding HEA formula from Refs.~\cite{Lukyanov2004a,Shukla2003}. In
Refs.~\cite{Farag2014a,Farag2014} a surface term to the ImOP was
added to improve the agreement with the data at lower energies.

In our present work, as well as in our previous works considering
processes with exotic He and Li isotopes
\cite{Lukyanov2007,Lukyanov2010,Lukyanov2009,Lukyanov2013}, we use
microscopically calculated OPs within the hybrid model
\cite{Lukyanov2004a}. In the latter the ReOP is calculated by a
folding of a nuclear density and the effective NN potentials
\cite{Khoa2000} (see also \cite{Lukyanov2007a}) and includes both
direct and exchange parts. The ImOP is obtained within the HEA
model \cite{Glauber, Sitenko}. There are only two or three fitting
parameters in the hybrid model that are related to the depths of
the ReOP, ImOP and the spin-orbit part of the OP. Along with some
phenomenological density distributions for He and Li isotopes we
have used in our works realistic microscopic density obtained
within the large-scale shell model (LSSM)
\cite{Karataglidis97,Karataglidis2000}. In the present work
devoted to processes with $^{10,11}$Be nuclei we use the density
distribution for $^{10}$Be obtained within the QMC model
\cite{Pieper2002,Pieper2002a} and also the densities of $^{10}$Be
and $^{11}$Be obtained within the GCM \cite{Descouvemont97}.

The main aim of our work is twofold. First, we study the elastic
scattering of the neutron-rich exotic $^{10}$Be and $^{11}$Be
nuclei on protons and nuclei at energies $E<100$ MeV/nucleon using
microscopically calculated in our work real and imaginary parts of
the optical potentials. Second, we estimate important
characteristics of the reactions with $^{11}$Be, such as the
breakup cross sections and momentum distributions of fragments in
breakup processes. To this end we use the model in which $^{11}$Be
consists of a core of $^{10}$Be and a halo formed by a motion of a
neutron in its periphery (e.g.,
Refs.~\cite{Fukuda91,Fukuda2004,Gravo2010}). The latter model is
justified by the small separation energy $S_{n}=504 \pm 6$ KeV of
a neutron from the ground $s_{1/2}$ state of $^{11}$Be
\cite{Sagava93} and on the observed quite large total interaction
cross sections of $^{11}$Be with target nuclei caused by the main
contribution from the breakup of $^{11}$Be on $^{10}$Be and a
neutron. The important role of the periphery is confirmed also by
the experiments on scattering of $^{11}$Be on the heavy nucleus of
$^{208}$Pb \cite{Nakamura94}, where the prevailing mechanism is
the direct breakup due to the long-range Coulomb force of the
nucleus. Also we should mention the important observation of the
narrow peak in the momentum distribution of the $^{10}$Be
fragments at the breakup of $^{11}$Be scattering on the $^{12}$C
nucleus \cite{Kelley95}, that is, as mentioned above, a
consequence of the large extension of the wave function of the
relative motion in the $^{10}$Be$+n$ system related to the small
neutron separation energy. By means of such a cluster model of
$^{11}$Be one can calculate the OPs for scattering of $^{11}$Be on
protons or nuclear targets. To this end one should use the known
$n$+$p$ potential and calculate using the microscopic model the
optical potentials of $^{10}$Be+$p$ (or $^{10}$Be+A and the $n$+A
potentials). Then the sum of these potentials are folded with a
density probability of the relative motion of the core $^{10}$Be
and the neutron. Also, in the framework of this cluster model one
can calculate the momentum distribution of $^{10}$Be fragments
from the breakup reactions $^{11}$Be+$^{9}$Be,
$^{11}$Be+$^{93}$Nb, $^{11}$Be+$^{181}$Ta, and $^{11}$Be+$^{238}$U
for which experimental data are available.

The structure of the paper is as follows. The theoretical scheme
to calculate microscopically within the hybrid model the ReOP,
ImOP, the spin-orbit part of the OP, the surface component of OP,
as well as the results of the calculations of the elastic
scattering cross sections of $^{10,11}$Be+$p$ and
$^{10,11}$Be+$^{12}$C are presented in Sect.~\ref{s:elastic}. The
basic expressions to estimate the breakup of $^{11}$Be and to
calculate the cross sections and the fragment momentum
distributions of $^{10}$Be in the diffraction and stripping
processes of $^{11}$Be on $^{9}$Be, $^{93}$Nb, $^{181}$Ta, and
$^{238}$U are given in Sect.~\ref{s:break}. The summary and
conclusions of the work are included in Sect.~\ref{s:summary}.

\section{Elastic scattering of $^{10,11}$Be on protons and $^{12}$C at $E<100$ MeV/nucleon}
\label{s:elastic}

\subsection{Hybrid model of the microscopic optical potential}

In the present work we calculate the microscopic OP that contains
the volume real ($V^{F}$) and imaginary parts (W), and the
spin-orbit interaction ($V^{ls}$). This OP is used for
calculations of elastic scattering differential cross sections. We
introduce a set of weighting coefficients $N_{R}$, $N_{I}$,
$N_{R}^{ls}$ and $N_{I}^{ls}$ that are related to the depths of
the corresponding parts of the OP and are obtained by a fitting
procedure to the available experimental data. Details of the
constructing of the OP are given in
Refs.\cite{Satchler79,Khoa1993,Khoa2000,Lukyanov2007a}. The OP has
the form:
\begin{widetext}
\begin{equation}\label{eq:1}
U(r)=N_R V^{F}(r) + iN_IW(r)- 2\lambda^{2}_{\pi}\left[ N_R^{ls}
V^{ls}_R\frac{1}{r} \frac{df_{R}(r)}{dr}+ i N_I^{ls} W_I^{ls}
\frac{1}{r} \frac{df_{I}(r)}{dr} \right]({\vec l}\cdot{\vec s}),
\end{equation}
\end{widetext}
where $2\lambda^{2}_{\pi}= 4$ fm$^{2}$ with the squared pion
Compton wave length $\lambda^{2}_{\pi}= 2$ fm$^{2}$. Let us denote
the values of the ReOP and ImOP at $r=0$ by $V_R(\equiv
V^{F}(r=0))$ and $W_I(\equiv W{(r=0)})$. We note that the
spin-orbit part of the OP contains real and imaginary terms with
the parameters $V^{ls}_{R}$ and $W^{ls}_{I}$ related to $V_{R}$
and $W_{I}$ by the $V^{ls}_R=V_R/4$ and $W^{ls}_I=W_I/4$,
correspondingly. Here $V_{R}$ and $W_{I}$ (and $V^{ls}_R$ and
$W^{ls}_I)$ have to be negative. The ReOP $V^{F}(r)$ is a sum of
isoscalar ($V^{F}_{IS}$) and isovector ($V^{F}_{IV}$) components
and each of them has its direct ($V^{D}_{IS}$ and $V^{D}_{IV}$)
and exchanged ($V^{EX}_{IS}$ and $V^{EX}_{IV}$) parts.

The isoscalar component has the form
\begin{widetext}
\begin{equation}\label{eq:2}
V_{IS}^{F}(r)= V_{IS}^{D}(r)+V_{IS}^{EX}(r)=\int d^{3}{\bf r}_p
d^{3}{\bf r}_t \{ \rho_p({\bf r}_p) \rho_t({\bf r}_t)
v_{NN}^D({\bf s}) + \rho_p({\bf r}_p,{\bf r}_p + {\bf s})
\rho_t({\bf r}_t,{\bf r}_t - {\bf s}) v_{NN}^{EX}({\bf s})
\exp[\imath {\bf K}(r) {\bf s}/M ] \},
\end{equation}
\end{widetext}
where ${\bf s}={\bf r}+{\bf r}_t-{\bf r}_p$ is the vector between
two nucleons, one of which belongs to the projectile and another
one to the target nucleus.

In the first term of the right-hand side of Eq.~(\ref{eq:2}) the
densities of the incident particle $\rho_{p}$ and the target
nucleus $\rho_{t}$  are sums of the proton and neutron densities.
In the second term $\rho_{p}$ and $\rho_{t}$ are the corresponding
one-body density matrices. In our work we use for them the
approximations for the knock-on exchange term of the folded
potential from Refs.~\cite{Campy78,Negele72} (see also
\cite{Lukyanov2007,Lukyanov2009}). In Eq.~(\ref{eq:2}) ${\bf
K}(r)$ is the local momentum of the nucleus-nucleus relative
motion and $v_{NN}^D$ and $v_{NN}^{EX}$ are the direct and
exchange effective NN potentials. They contain an energy
dependence usually taken in the form $g(E)=1- 0.003 E$ and a
density dependence with the form for the CDM3Y6 effective Paris
potential \cite{Khoa2000}
\begin{equation}\label{eq:3}
F(\rho)=C\left [1+\alpha e^{-\beta\rho({\bf r})}-\gamma\rho({\bf
r})\right ]
\end{equation}
with  $C$=0.2658, $\alpha$=3.8033, $\beta$=1.4099 fm$^3$, and
$\gamma$=4.0 fm$^3$. The effective NN interactions $v_{NN}^D$ and
$v_{NN}^{EX}$ have their isoscalar and isovector components in the
form of M3Y interaction obtained within $g$-matrix calculations
using the Paris NN potential \cite{Khoa1993,Khoa2000}. The
isovector components $V_{IV}^{F}$ of the ReOP can be obtained by
exchanging in Eq.~(\ref{eq:2}) the sum of the proton and neutron
densities in $\rho_{p(t)}$ by their difference and using the
isovector parts of the effective NN interaction. In the case of
the proton scattering on nuclei Eq.~(\ref{eq:2}) contains only the
density of the target nucleus.

The ImOP can be chosen either to be in the form of the
microscopically calculated $V^{F}$ ($W=V^{F}$) or in the form
$W^{H}$ obtained in Ref.~\cite{Lukyanov2004a,Shukla2003} within
the HEA of the scattering theory \cite{Glauber,Sitenko}:
\begin{eqnarray}
W^H(r)=-\frac{\bar{\sigma}_{N}}{2\pi^{2}}\frac{E}{k}\int_{0}^{\infty}
j_0(kr)\rho_p(q)\rho_t(q)f_{N}(q)q^2 dq .
\label{eq:4}
\end{eqnarray}
In Eq.~(\ref{eq:4}) $\rho(q)$ are the corresponding formfactors of
the nuclear densities, $f_N(q)$ is the amplitude of the NN
scattering and $\bar\sigma_{N}$ is the averaged over the isospin
of the nucleus total NN scattering cross section that depends on
the energy. The parametrization of the latter dependence can be
seen, e.g., in Refs.~\cite{Charagi90,Lukyanov2007}. We note that
to obtain the HEA OP (with its imaginary part $W^{H}$ in
Eq.(~\ref{eq:4})) one can use the definition of the eikonal phase
as an integral of the nucleon-nucleus potential over the
trajectory of the straight-line propagation and has to compare it
with the corresponding Glauber expression for the phase in the
optical limit approximation. In the suggested scheme we use the
nuclear densities and NN cross sections known from other sources
and also the already used NN potentials and amplitudes. In this
way, the only free parameters in our approach are the parameters
$N$s that renormalize the depths of the OPs components. In the
spin-orbit parts of the OP the functions $f_i$(r) ($i=R,I$)
correspond to WS forms of the potentials with parameters of the
real and imaginary parts $V_R$, $W_I$, $R_i$, $a_i$
[$f_R(r,R_R,a_R)$ and $f_I(r,R_I,a_I)$], as they are used in the
DWUCK4 code \cite{DWUCK} and applied for numerical calculations.
We determine the values of these parameters by fitting the WS
potentials to the microscopically calculated potentials $V^F$(r)
and W(r).

\subsection{Results of calculations of elastic scattering cross sections}

\begin{figure*}
\includegraphics[width=0.7\linewidth]{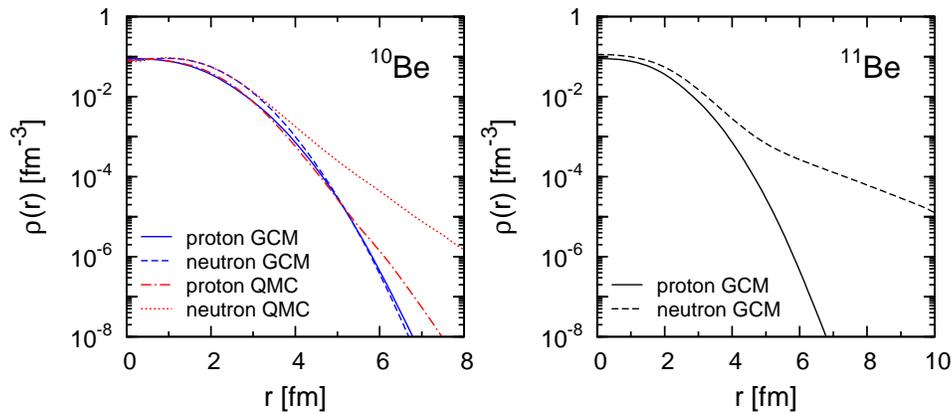}
\caption{(Color online) Point-proton (normalized to $Z=4$) and
point-neutron (normalized to $N=6$ and $N=7$, respectively)
densities of $^{10}$Be and $^{11}$Be obtained in the GCM
\protect\cite{Descouvemont97} and in the QMC method
\protect\cite{Pieper2002,Pieper2002a}.
\label{fig1}}
\end{figure*}

In the calculations of the microscopic OPs for the scattering of
$^{10,11}$Be on protons and nuclei we used realistic density
distributions of $^{10}$Be calculated within the quantum Monte
Carlo model \cite{Pieper2002,Pieper2002a} and of $^{10,11}$Be from
the generator coordinate method \cite{Descouvemont97}. In general,
the QMC methods include both variational and Green's function
Monte Carlo methods. In our case within the QMC method the proton
and neutron densities of $^{10}$Be have been computed with the
AV18+IL7 Hamiltonian \cite{Pieper2002a}. As far as the GCM
densities are concerned, in Ref.~\cite{Descouvemont97} the
$^{10}$Be wave functions are defined in the harmonic oscillator
model with all $p$-shell configurations. The $^{11}$Be wave
functions are described in terms of cluster wave functions,
relative to $^{10}$Be and to the external neutron. Thus, both
microscopic densities effectively account for the non-ordinary
nuclear structure peculiarities of $^{10,11}$Be
\cite{Shrivastava2004,Crespo2011} and their use is physically
justified. The QMC and GCM densities are given in Fig.~\ref{fig1}.
It can be seen that they have been calculated with enough accuracy
up to distances much larger than the nuclear radius. In both
methods the densities of $^{10}$Be occur quite similar up to
$r\sim3.5$ fm and a difference between them is seen in their
asymptotics. In the calculations of the OPs for
$^{10,11}$Be+$^{12}$C the density of $^{12}$C was taken in
symmetrized Fermi form with radius and diffuseness parameters
$c=3.593$ fm and $a=0.493$ fm, respectively \cite{Burov77}. The
results of the calculations are compared with the available
experimental data. All calculations of elastic scattering using
the obtained OPs are performed by using the DWUCK4 code
\cite{DWUCK}.

\subsubsection{Elastic scattering cross sections of
$^{10,11}$Be+$p$}

On the basis on the scheme presented in subsection II.A. we
calculated the elastic scattering cross sections of
$^{10,11}$Be+$p$ and compared them with the available experimental
data.

It is accepted that the elastic scattering of light nuclei is
rather sensitive to their periphery, where transfer and breakup
processes also take place. Therefore, investigating the elastic
scattering, one must bear in mind that virtual non-elastic
contributions can also take part in the process. It has been
pointed out in our previous papers
\cite{Lukyanov2009,Lukyanov2010}, as well as in
Refs.~\cite{Farag2012,Farag2014,Farag2014a}, that the inclusion of
a surface imaginary term to the OP [Eq.~(\ref{eq:1})] leads to a
better agreement with the experimental data. As known, this
contribution can be considered as the so-called dynamical
polarization potential, which allows one to simulate the surface
effects caused by the latter. In fact, the imaginary part of the
$ls$ term in our OP [see Eq.~(\ref{eq:1})] plays effectively this
role. However, sometimes one needs to increase the absorption in
the surface region and thus, one adds a derivative of the ImOP
(surface term):
\begin{equation}\label{eq:5a}
W^{sf}(r)=-i N_{I}^{sf}r\frac{dW(r)}{dr},
\end{equation}
where $N_{I}^{sf}$ is also a fitting parameter.

\begin{figure*}
\includegraphics[width=0.75\linewidth]{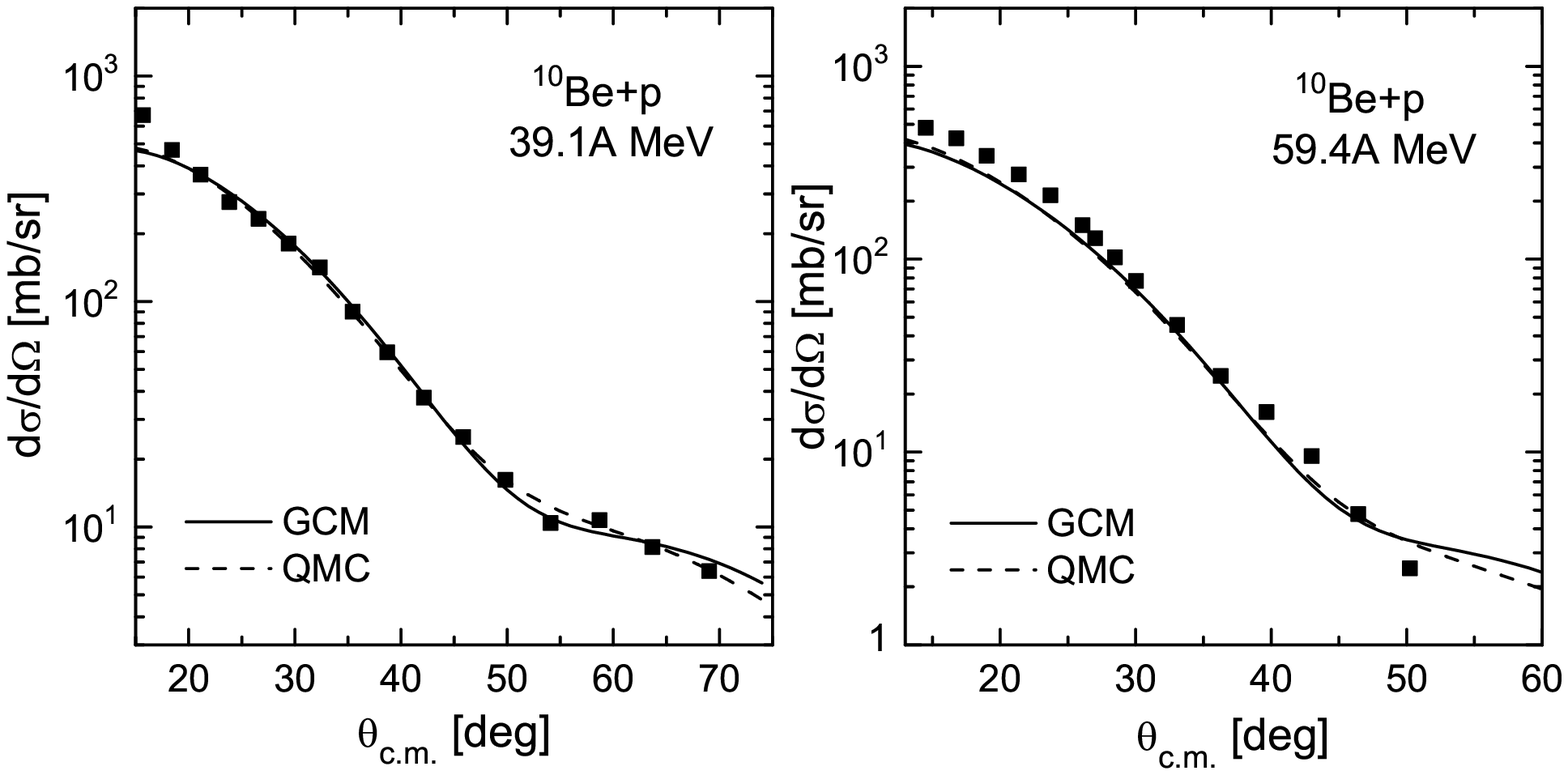}
\includegraphics[width=0.75\linewidth]{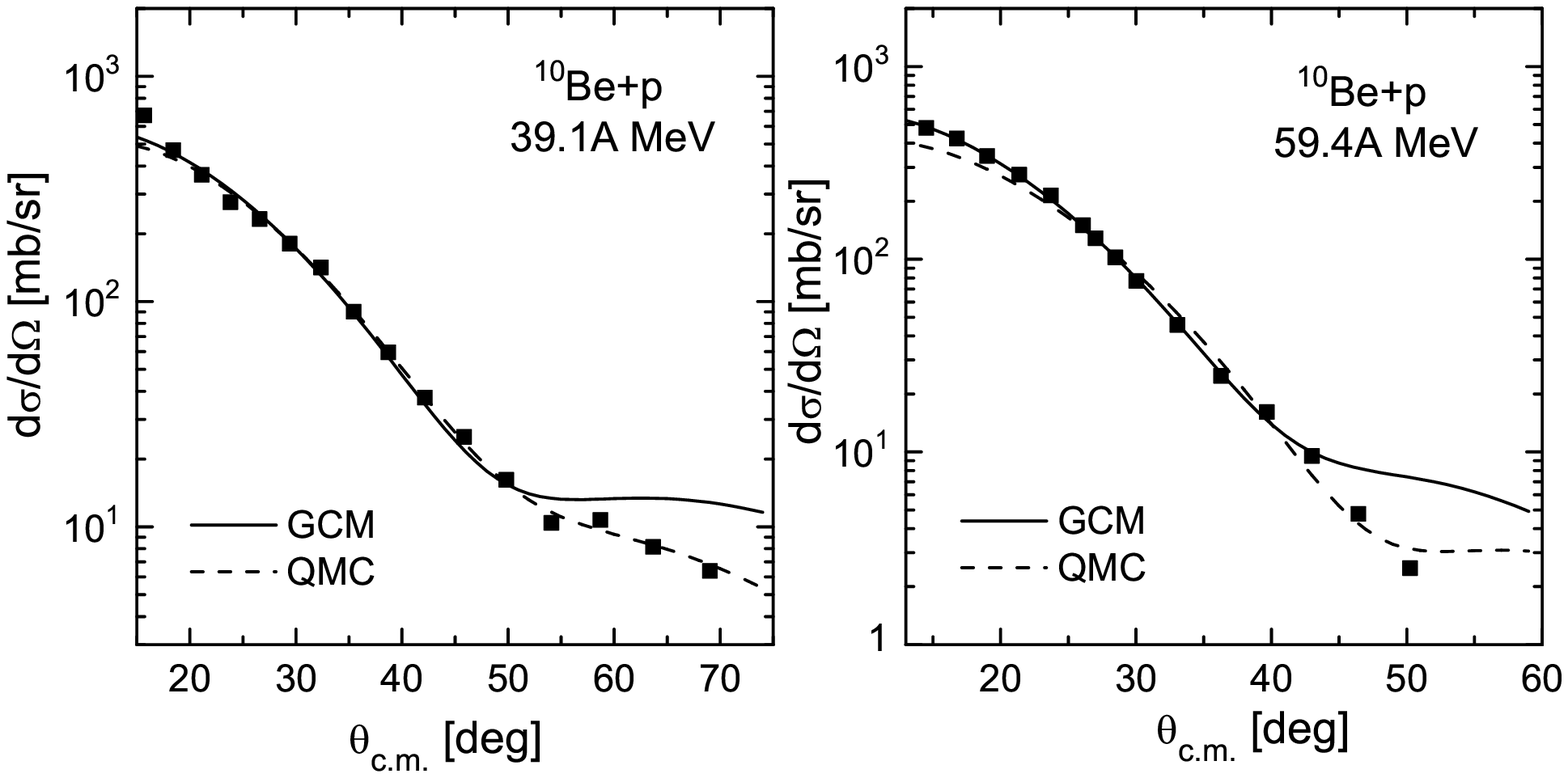}
\includegraphics[width=0.75\linewidth]{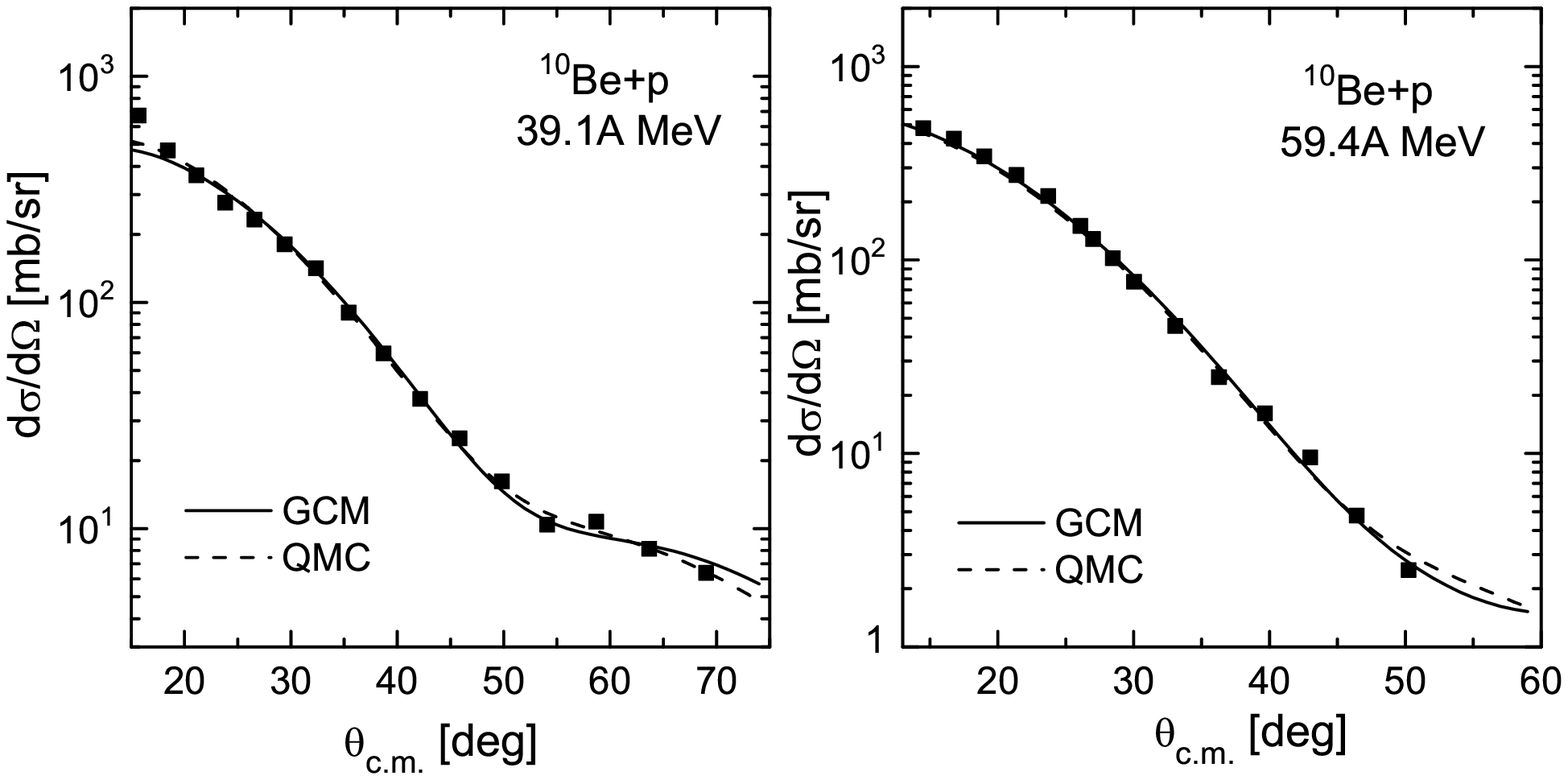}
\caption{$^{10}$Be+$p$ elastic scattering cross sections. Upper
panel: without $ls$ term; middle panel: with $ls$ term; bottom
panel: with both $ls$ and surface terms. Solid lines: calculations
with GCM density of $^{10}$Be; dashed lines: calculations with QMC
density of $^{10}$Be. Experimental data for 39.1 MeV/nucleon and
59.4 MeV/nucleon are taken from Refs.~\protect\cite{Lapoux2008}
and \protect\cite{Cortina-Gil97}, respectively.
\label{fig2}}
\end{figure*}

\begin{figure*}
\includegraphics[width=0.75\linewidth]{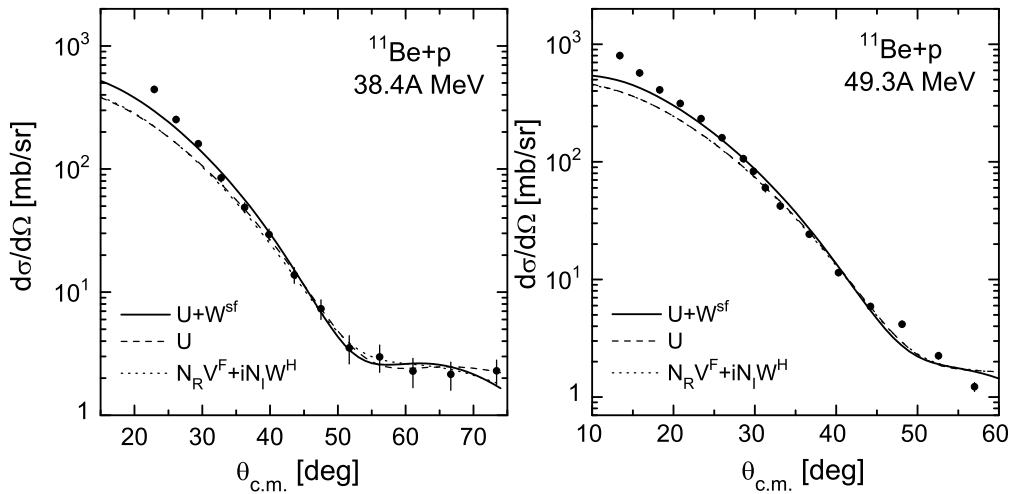}
\caption{$^{11}$Be+$p$ elastic scattering cross sections.
Calculations are performed with GCM density of $^{11}$Be. Solid
line: OP with both $ls$ and surface terms [Eqs.~(\ref{eq:1}) and
(\ref{eq:5a})]; dashed line: OP with $ls$ term [Eq.~(\ref{eq:1})];
dotted line: the volume part of OP from Eq.~(\ref{eq:1}).
Experimental data for 38.4 MeV/nucleon and 49.3 MeV/nucleon are
taken from Refs.~\protect\cite{Lapoux2008} and
\protect\cite{Cortina-Gil97}, respectively.
\label{fig3}}
\end{figure*}

The results for the elastic $^{10}$Be+$p$ and $^{11}$Be+$p$
scattering cross sections are given in Figs.~\ref{fig2} and
\ref{fig3}, respectively, and compared with the data at energies
39.1 MeV/nucleon \cite{Lapoux2008} and $59.4$ MeV/nucleon
\cite{Cortina-Gil97} for $^{10}$Be and 38.4 MeV/nucleon
\cite{Lapoux2008} and $49.3$ MeV/nucleon \cite{Cortina-Gil97} (see
also \cite{Lukyanov2014}) for $^{11}$Be. In general, our analysis
points out that more successful results are obtained in the case
when the ImOP is taken from HEA: $W(r)=W^{H}(r)$
[Eq.~(\ref{eq:4})]. We note that in the fitting procedure of the
theoretical results to the data for elastic scattering cross
sections for $^{10,11}$Be+$p$ (and also for $^{10,11}$Be+$^{12}$C)
there arises an ambiguity in the choice of the optimal curve among
many of them that are close to the experimental data. Due to this
we impose a physical constraint, namely choosing those ReOP and
ImOP that give volume integrals which have a correct dependence on
the energy. The volume integrals have the forms
\begin{equation}\label{eq:5aa}
J_V(E)=- \frac{4\pi}{A_p A_t}\int dr r^2 [N_{R}V^{F}(r)],
\end{equation}
\begin{equation}\label{eq:5b}
J_W^{(a)}(E)=- \frac{4\pi}{A_p A_t}\int dr r^2 [N_{I}W(r)],
\end{equation}
\begin{equation}\label{eq:5c}
J_W^{(b)}(E)=- \frac{4\pi}{A_p A_t}\int dr r^2 \left
[N_{I}W(r)-N_I^{sf}r\frac{dW(r)}{dr}\right ],
\end{equation}
where $A_p$ and $A_t$ are the mass numbers of the projectile and
the target, respectively. In Eq.~(\ref{eq:5c}) we added also the
integral over the surface term of the OP (\ref{eq:5a}). It is
known \cite{Romanovsky} that the volume integrals (their absolute
values) for the ReOP decrease with the increase of the energy,
while for the ImOP they increase up to a plateau and then
decrease. The values of the $N$s parameters from the fitting
procedure and after imposing the mentioned constraint are given in
Table~\ref{tab1}. It can be seen that the tendency (the decrease
of $J_V$ and the increase of $J_W$) is generally confirmed.

The calculated differential cross sections of $^{10}$Be+$p$
elastic scattering at energies $39.1$ MeV/nucleon and $59.4$
MeV/nucleon are presented in Fig.~\ref{fig2}. First, it is seen
from the upper panel that the inclusion of only the volume OP is
not enough to reproduce reasonably well the data in the small
angles region. Then, after adding the spin-orbit component to the
OP the agreement with the data becomes better, in particular for
the angular distributions calculated using the GCM density at
energies $39.1$ MeV/nucleon and $59.4$ MeV/nucleon for angles less
than 20$^{\circ}$ and 30$^{\circ}$, correspondingly, as
illustrated in the middle panel of Fig.~\ref{fig2}. However, a
discrepancy at larger angles remains. At the same time for the
cross sections with the account for the $ls$ interaction and using
the QMC density we obtain fairly good agreement with the data at
both energies and only a small discrepancy is seen at small angles
at energy $59.4$ MeV/nucleon. Further improvement is achieved when
both $ls$- and surface terms are included in the calculations. In
this case, as it can be seen from the bottom panel of
Fig.~\ref{fig2}, the discrepancy between the differential cross
sections for the GCM density and the experimental data at larger
angles is strongly reduced.

\begin{table*}
\caption{The renormalization parameters $N_{R}$, $N_{I}$,
$N_{R}^{ls}$, $N_{I}^{ls}$, and $N_{I}^{sf}$, the total reaction
cross sections $\sigma_{R}$ (in mb), and the volume integrals
$J_{V}$, $J_{W}^{(a)}$, and $J_{W}^{(b)}$ (in MeV.fm$^{3}$) as
functions of the energy $E$=39.1 and 59.4 MeV/nucleon for the
$^{10}$Be+$p$ and $E$=38.4 and 49.3 MeV/nucleon for the
$^{11}$Be+$p$ elastic scattering} \label{tab1}
\begin{center}
\begin{tabular}{cccccccccccc}
\hline \hline \noalign{\smallskip}
Nucleus & Model & $E$ & $N_R$ & $N_I$ & $N_{R}^{ls}$ & $N_{I}^{ls}$ & $N_{I}^{sf}$ & $\sigma_R$ & $J_V$ & $J_W^{(a)}$ & $J_W^{(b)}$\\
\noalign{\smallskip}\hline\noalign{\smallskip}
$^{10}$Be         & GCM & 39.1 & 0.983 & 0.267 & 0.000 & 0.000 & 0.000 & 292.12 & 389.408 & 116.600 & 116.600  \\
without $ls$ and  & QMC &      & 1.153 & 0.295 & 0.000 & 0.000 & 0.000 & 311.36 & 411.344 & 130.806 & 130.806  \\
surface terms     & GCM & 59.4 & 1.001 & 0.802 & 0.000 & 0.000 & 0.000 & 341.18 & 333.739 & 263.540 & 263.540  \\
                  & QMC &      & 1.188 & 0.856 & 0.000 & 0.000 & 0.000 & 356.98 & 354.606 & 283.464 & 283.464  \\
\noalign{\smallskip}\hline\noalign{\smallskip}
$^{10}$Be         & GCM & 39.1 & 1.493 & 0.492 & 1.000 & 0.476 & 0.000 & 372.50 & 591.440 & 216.480 & 216.480  \\
with $ls$ and     & QMC &      & 1.163 & 0.318 & 0.557 & 0.000 & 0.000 & 323.96 & 414.911 & 141.004 & 141.004  \\
without surface   & GCM & 59.4 & 1.294 & 0.804 & 0.190 & 0.000 & 0.000 & 355.29 & 431.427 & 264.197 & 264.197  \\
terms             & QMC &      & 1.014 & 0.527 & 0.940 & 0.000 & 0.000 & 287.68 & 302.669 & 174.516 & 174.516  \\
\noalign{\smallskip}\hline\noalign{\smallskip}
$^{10}$Be         & GCM & 39.1 & 0.995 & 0.266 & 0.095 & 0.082 & 0.004 & 298.65 & 394.161 & 117.040 & 122.321  \\
with $ls$ and     & QMC &      & 1.194 & 0.260 & 0.075 & 0.025 & 0.018 & 333.71 & 425.971 & 115.286 & 139.235  \\
surface terms     & GCM & 59.4 & 0.970 & 0.000 & 0.365 & 1.000 & 0.373 & 400.26 & 323.404 &   0.000 & 367.802  \\
                  & QMC &      & 1.043 & 0.281 & 0.000 & 1.000 & 0.270 & 389.27 & 311.325 &  93.053 & 361.343  \\
\noalign{\smallskip}\hline\noalign{\smallskip}
$^{11}$Be         & GCM & 38.4 & 0.824 & 0.659 & 0.000 & 0.000 & 0.000 & 459.05 & 339.388 & 293.493 & 293.493  \\
without $ls$ and  &     & 49.3 & 0.793 & 0.805 & 0.000 & 0.000 & 0.000 & 423.52 & 296.301 & 301.184 & 301.184  \\
surface terms     &     &      &       &       &       &       &       &        &         &         &          \\
\noalign{\smallskip}\hline\noalign{\smallskip}
$^{11}$Be         & GCM & 38.4 & 0.787 & 0.799 & 0.000 & 0.507 & 0.000 & 458.63 & 324.148 & 355.844 & 355.844  \\
with $ls$ and     &     & 49.3 & 0.793 & 0.867 & 0.123 & 0.316 & 0.000 & 426.85 & 296.301 & 301.184 & 301.184  \\
without surface   &     &      &       &       &       &       &       &        &         &         &          \\
terms             &     &      &       &       &       &       &       &        &         &         &          \\
\noalign{\smallskip}\hline\noalign{\smallskip}
$^{11}$Be with    & GCM & 38.4 & 0.849 & 0.106 & 0.102 & 0.380 & 0.152 & 493.01 & 349.685 &  47.208 & 269.903  \\
$ls$ and surface  &     & 49.3 & 0.801 & 0.000 & 0.213 & 0.394 & 0.200 & 436.46 & 299.280 &   0.000 & 246.162  \\
terms             &     &      &       &       &       &&&&&&                                                  \\
\noalign{\smallskip}\hline \hline
\end{tabular}
\end{center}
\end{table*}

In general, the account for the spin-orbit term in the volume OP
gives a trend of an increase of the  cross sections at larger
angles, that seems to be related with the change of the form of
the total OP at its periphery. If we evaluate the quantities of
the two densities of $^{10}$Be on the basis of the  values of the
parameter $N_R$ (comparing which ones are closer to unity), our
conclusion is that in the calculations without $ls$ interaction
the GCM density works better, while in the case with $ls$ term in
the OP the QMC density gives better results. A fair agreement
between the calculated $^{10}$Be+$p$ angular distributions and the
experimental data is obtained only when both $ls$- and surface
contributions to the OP are included.

In Fig.~\ref{fig3} are given and compared with the empirical data
elastic cross sections for the scattering of $^{11}$Be on protons
at energies 38.4 and 49.3 MeV/nucleon applying the fitting
procedure for the parameters $N$s. All of them are calculated
using GCM density of $^{11}$Be. The different curves drawn in
Fig.~\ref{fig3} correspond to those given in Fig.~\ref{fig2} with
accounting for different contributions to the OP. One can see a
discrepancy at small angles ($\theta<30^{\circ}$) that seems to be
related to the contributions from the surface region of
interactions, where breakup processes play an important role.
Similarly to the results for the $^{10}$Be+$p$ elastic scattering
cross sections (see Fig.~\ref{fig2}), the account for both
spin-orbit and surface terms to the OP leads to a better agreement
with the $^{11}$Be+$p$ data in the region of small angles. In
Table~\ref{tab1} are given the corresponding values of the
parameters $N_R$ and $N_I$ whose values deviate from unity of
about $20-30\%$ that points out that the hybrid model for the O*P
can be used successfully in such calculations.

We would like to emphasize the fact that when considering the case
of the total OP [Eqs.~(\ref{eq:1}) and (\ref{eq:5a})], the values
of the parameters $N_I$ drop down sufficiently in comparison with
their values coming out from the two other cases. They are
compensated in the most cases by the non-zero values of
$N_R^{ls}$, $N_I^{ls}$, and $N_I^{sf}$ parameters. Here we would
like to note that the $ls$ term used in our calculations (with
both real and imaginary parts) plays the similar role as the
surface term applied in Ref.~\cite{Farag2014a}, where, however,
the imaginary $ls$ term is disregarded. From our analysis made for
the elastic scattering of $^{10}$Be and $^{11}$Be on protons we
conclude also that the surface imaginary part of the OP is less
necessary to fit the data of proton elastic scattering on the
stable nucleus $^{10}$Be, but it is important to give an agreement
with the proton elastic-scattering data of the halo nucleus
$^{11}$Be. This is mainly due to the specific halo structure of
the $^{11}$Be density distribution and its large rms radius.

\begin{figure*}
\includegraphics[width=0.7\linewidth]{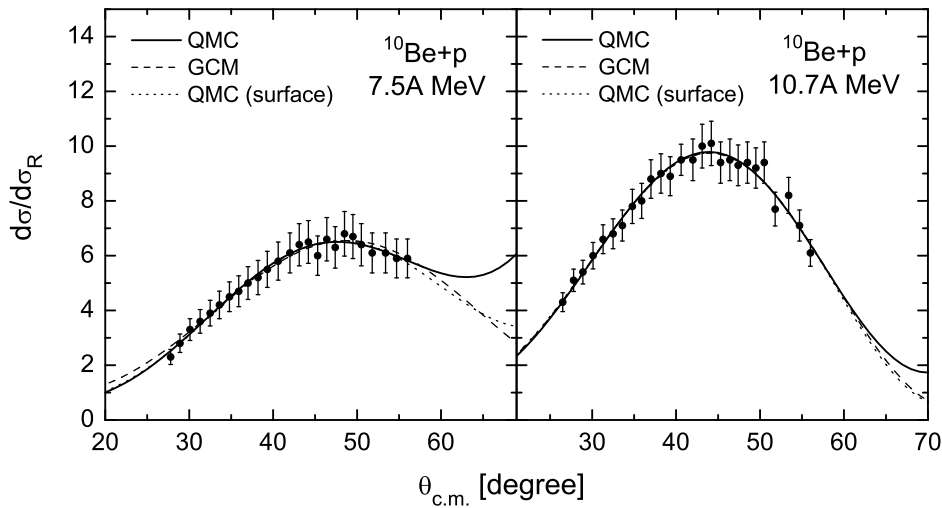}
\caption{$^{10}$Be+$p$ elastic scattering cross sections as a
ratio to Rutherford scattering at proton energies of 7.5 MeV (left
panel) and 10.7 MeV (right panel). The solid and dashed lines show
the results with QMC and GCM density of $^{10}$Be, respectively,
and with $ls$ term in OP. The dotted lines show the QMC results
obtained by accounting for both the $ls$- and surface terms in OP.
Experimental data are taken from Ref.~\protect\cite{Schmitt2013}.
\label{fig3a}}
\end{figure*}

For a more complete analysis of the elastic scattering cross
sections, we extend the incident energy region to lower energies
on the example of the scattering of $^{10}$Be on protons that has
been recently studied by Schmitt {\it et al.} \cite{Schmitt2013}.
Moreover, this could be a test of our hybrid model at low
energies. In Ref.~\cite{Schmitt2013} proton energies of 6, 7.5, 9,
and 10.7 MeV were selected to measure the elastic scattering cross
sections for protons with $^{10}$Be beams in inverse kinematics in
order to provide constraints on optical potentials for reaction
studies with light neutron-rich nuclei. The calculated results for
the differential cross sections, shown as a ratio to Rutherford
scattering, are given and compared with the data
\cite{Schmitt2013} in Fig.~\ref{fig3a} for energies of 7.5 and
10.7 MeV. The values of the $N$s parameters from the fitting
procedure and the corresponding total reaction cross sections and
volume integrals are listed in Table~\ref{tab1a}. The results
shown in Fig.~\ref{fig3a} when including in the calculations only
the $ls$ term demonstrate a fairly good agreement with the data.
The values of the parameters $N_{R}$ deduced from the fitting
procedure for both energies in the case of GCM density of
$^{10}$Be are quite large that indicates for the specific
peculiarities of the elastic scattering at low energies with
account for the spin-orbit term. We also calculated the
$^{10}$Be+$p$ elastic scattering cross sections at the same proton
energies taking into account the surface term [Eq.~(\ref{eq:5a})].
In this case, only the QMC density of $^{10}$Be was tested that
has been also used in Ref.~\cite{Farag2014a}, where the two other
energies of 6 and 9 MeV were considered. The results illustrate
that the inclusion of the surface contribution does not affect the
good agreement obtained without it. Here we note that in
Ref.~\cite{Schmitt2013} no single optical potential had been found
to reproduce well the proton elastic scattering data over this
range of energies. At the same time, it is seen from the left
panel of Fig.~\ref{fig3a} a deviation of the results of our model
with both densities beyond 55$^\circ$. Therefore, it would be
desirable to measure the elastic channel in this angular range to
constrain the $p$--$^{10}$Be optical potential.

\begin{table*}
\caption{The renormalization parameters $N_{R}$, $N_{I}$,
$N_{R}^{ls}$, $N_{I}^{ls}$, and $N_{I}^{sf}$, the total reaction
cross sections $\sigma_{R}$ (in mb), and the volume integrals
$J_{V}$, $J_{W}^{(a)}$, and $J_{W}^{(b)}$ (in MeV.fm$^{3}$) as
functions of the proton energy $E$=7.5 and 10.7 MeV for the
$^{10}$Be+$p$ elastic scattering} \label{tab1a}
\begin{center}
\begin{tabular}{cccccccccccc}
\hline \hline \noalign{\smallskip}
Nucleus & Model & $E$ & $N_R$ & $N_I$ & $N_{R}^{ls}$ & $N_{I}^{ls}$ & $N_{I}^{sf}$ & $\sigma_R$ & $J_V$ & $J_W^{(a)}$ & $J_W^{(b)}$\\
\noalign{\smallskip}\hline\noalign{\smallskip}
$^{10}$Be         & GCM & 7.5  & 2.287 & 0.473 & 0.000 & 0.425 & 0.000 & 906.19 & 1215.283 & 527.749  & 527.749  \\
with $ls$ and     & QMC &      & 1.244 & 0.056 & 0.065 & 0.103 & 0.000 & 330.03 & 603.634  & 62.966   & 62.966   \\
without surface   & GCM & 10.7 & 2.232 & 1.129 & 0.000 & 0.759 & 0.000 & 804.23 & 1144.742 & 1151.009 & 1151.009 \\
terms             & QMC &      & 1.915 & 0.247 & 0.963 & 0.307 & 0.000 & 722.54 & 895.179  & 253.766  & 253.766  \\
\noalign{\smallskip}\hline\noalign{\smallskip}
$^{10}$Be         & QMC & 7.5  & 1.483 & 0.000 & 0.442 & 0.208 & 0.044 & 306.28 & 719.605  & 0.000    & 148.453  \\
with $ls$ and     & QMC & 10.7 & 1.354 & 0.098 & 0.178 & 1.000 & 0.193 & 636.50 & 632.936  & 100.685  & 695.676  \\
surface terms     &     &      &       &       &       &&&&&&                                                    \\
\noalign{\smallskip}\hline \hline
\end{tabular}
\end{center}
\end{table*}

\subsubsection{Elastic scattering cross sections of $^{10,11}$Be+$^{12}$C}

\begin{figure*}
\includegraphics[width=0.4\linewidth]{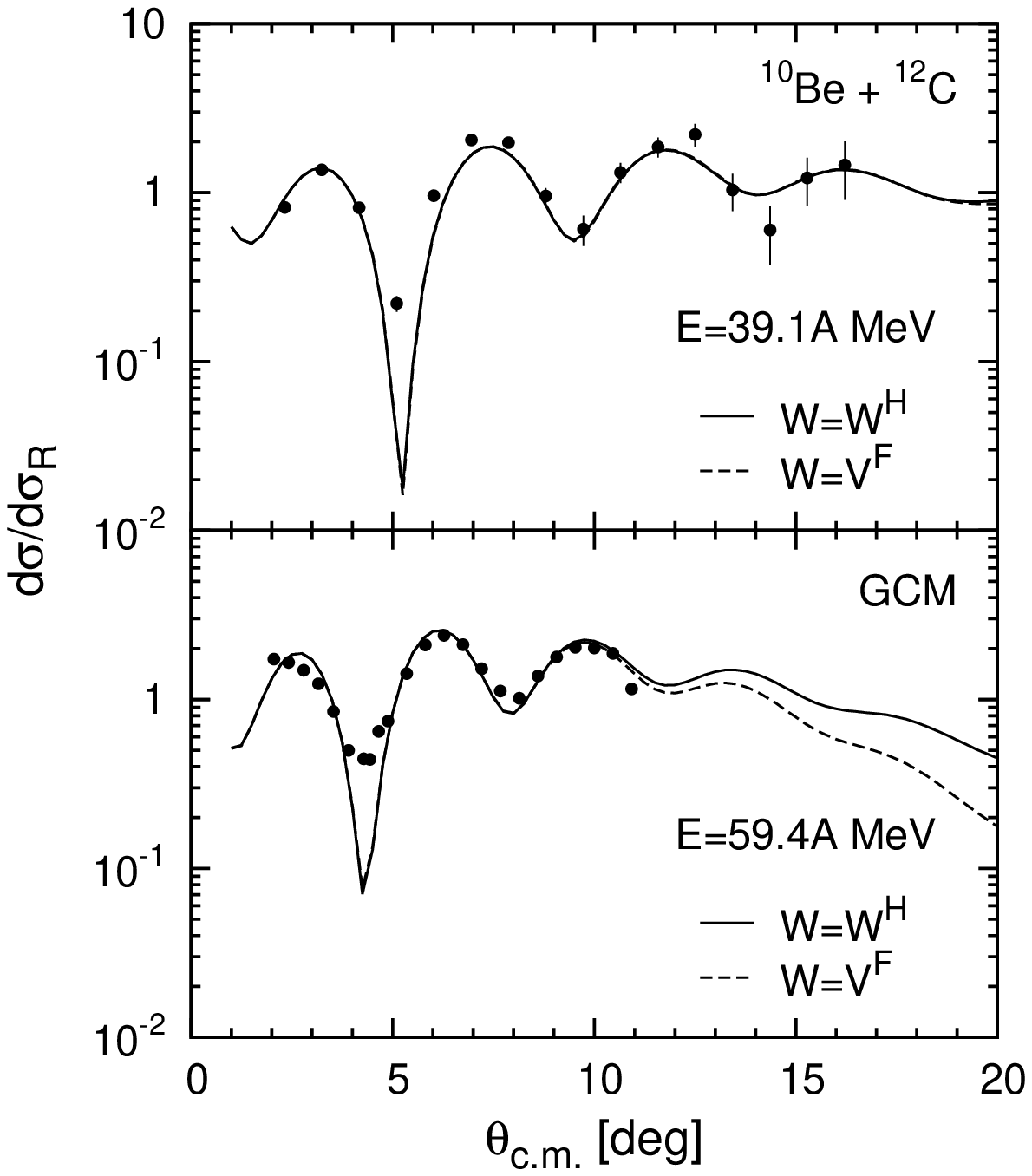}
\includegraphics[width=0.4\linewidth]{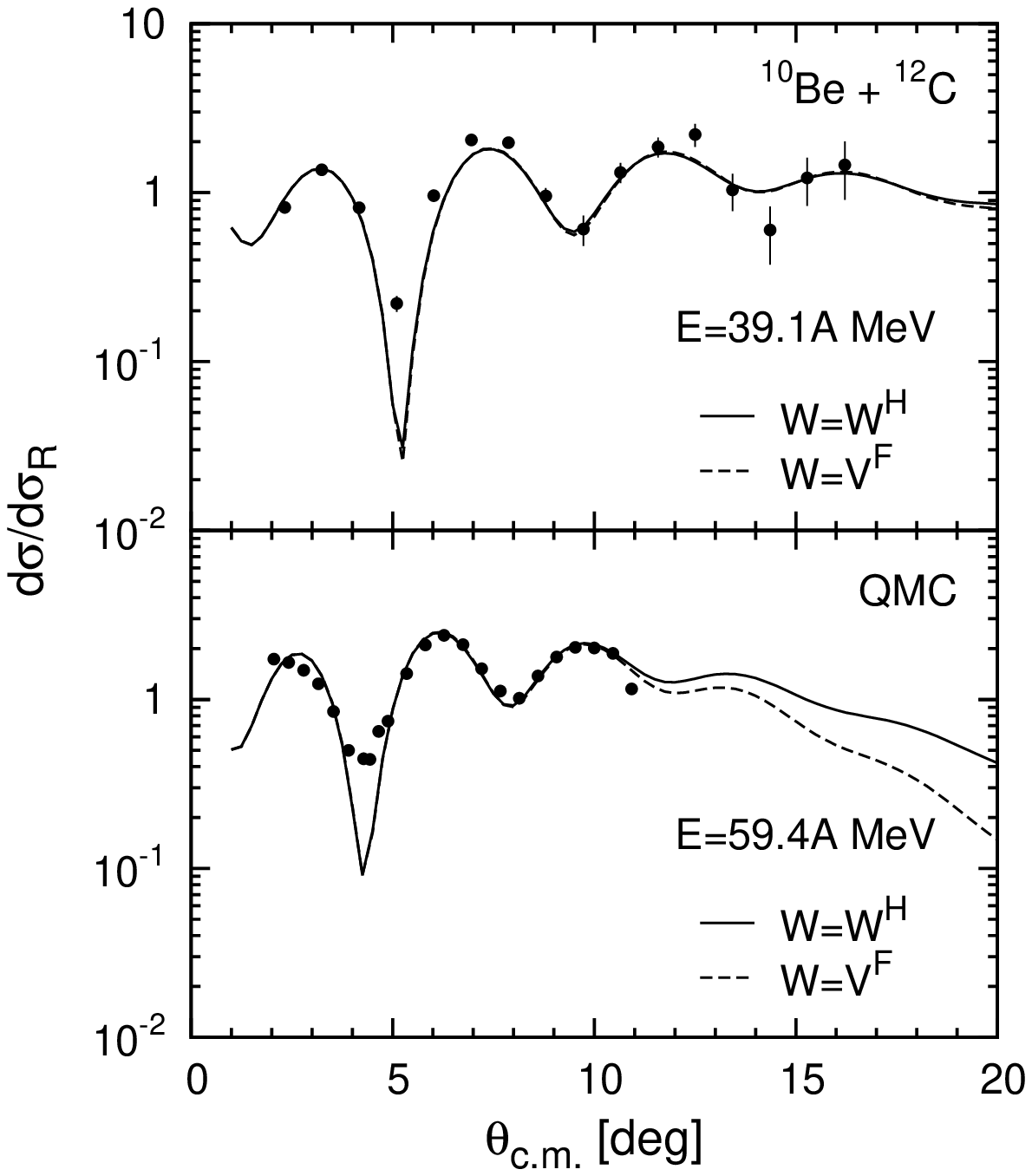}
\caption{$^{10}$Be+$^{12}$C elastic scattering cross sections.
Solid lines: $W=W^H$; dashed lines: $W=V^F$. Left panel:
calculations with GCM density of $^{10}$Be; right panel:
calculations with QMC density of $^{10}$Be. Experimental data for
39.1 MeV/nucleon and 59.4 MeV/nucleon are taken from
Refs.~\protect\cite{Lapoux2008} and \protect\cite{Cortina-Gil97},
respectively. \label{fig4}}
\end{figure*}

\begin{figure}
\includegraphics[width=0.9\linewidth]{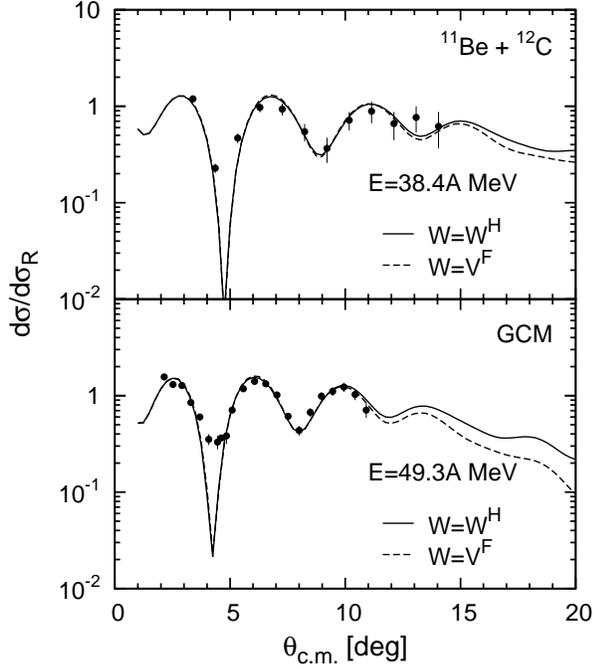}
\caption{$^{11}$Be+$^{12}$C elastic scattering cross sections.
Solid lines: $W=W^H$; dashed lines: $W=V^F$. For $^{11}$Be GCM
density was used. Experimental data for 38.4 MeV/nucleon and 49.3
MeV/nucleon are taken from Refs.~\protect\cite{Lapoux2008} and
\protect\cite{Cortina-Gil97}, respectively.
\label{fig5}}
\end{figure}

The calculated within the hybrid model elastic scattering cross
sections of $^{10,11}$Be+$^{12}$C (their ratios to the Rutherford
one) at the same energies, as for $^{10,11}$Be+$p$ scattering, are
given in Figs.~\ref{fig4} and \ref{fig5} and compared with the
experimental data (see also \cite{Lukyanov2014}). In comparison
with the case of $^{10,11}$Be+$p$, the experimental data
\cite{Cortina-Gil97,Lapoux2008} for the scattering on $^{12}$C
demonstrate more developed diffractional picture on the basis of
the stronger influence of the Coulomb field. It can be seen in
Fig.~\ref{fig4} that in both cases of calculations of OPs with QMC
or GCM densities the results are in a good agreement with the
available data. It is seen also from the figures that it is
difficult to determine the advantage of the use for the ImOP
$W=W^{H}$ or $W=V^{F}$, because the differences between the
theoretical results start at angles for which the experimental
data are not available. The values of the parameters $N_R$ and
$N_I$ (the depths  of ReOP and ImOP) are given in
Table~\ref{tab2}. From the comparison of these values, when GCM or
QMC densities are used, one can see that in the case of GCM
densities the values of the parameters are closer to unity. In
this way, we may conclude that as in the $^{10}$Be+$p$ case
without $ls$ term of OP, the GCM density can be considered as a
more realistic one.

\begin{table*}
\caption{The renormalization parameters $N_{R}$ and $N_{I}$, the
total reaction cross sections $\sigma_{R}$ (in mb), and the volume
integrals $J_{V}$ and $J_{W}$ (in MeV.fm$^{3}$) as functions of
the energy $E$=39.1 and 59.4 MeV/nucleon for the
$^{10}$Be+$^{12}$C and $E$=38.4 and 49.3 MeV/nucleon for the
$^{11}$Be+$^{12}$C elastic scattering} \label{tab2}
\begin{center}
\begin{tabular}{ccccccccc}
\hline \hline \noalign{\smallskip}
Nucleus & Model & $E$ & $W$ & $N_R$ & $N_I$ & $\sigma_R$ & $J_V$ & $J_W^{(a)}$ \\
\noalign{\smallskip}\hline\noalign{\smallskip}
$^{10}$Be    & GCM & 39.1 & $W^{H}$ & 0.939 & 0.708 & 104.539 & 255.156 & 283.037 \\
             &     &      & $V^{F}$ & 0.816 & 0.465 & 105.958 & 221.733 & 126.355 \\
             &     & 59.4 & $W^{H}$ & 1.013 & 1.010 & 101.052 & 238.122 & 302.581 \\
             &     &      & $V^{F}$ & 0.884 & 0.577 & 102.635 & 207.798 & 135.633 \\
\noalign{\smallskip}\hline\noalign{\smallskip}
$^{10}$Be & QMC & 39.1 & $W^{H}$ & 0.888 & 0.620 & 105.332 & 245.613 & 249.769 \\
          &     &      & $V^{F}$ & 0.782 & 0.434 & 106.878 & 216.294 & 120.041 \\
          &     & 59.4 & $W^{H}$ & 0.970 & 0.887 & 101.616 & 231.953 & 267.782 \\
          &     &      & $V^{F}$ & 0.849 & 0.534 & 103.035 & 203.019 & 127.694 \\
\noalign{\smallskip}\hline\noalign{\smallskip}
$^{11}$Be & GCM & 38.4 & $W^{H}$ & 0.769 & 0.711 & 127.123 & 216.879 & 287.235 \\
          &     &      & $V^{F}$ & 0.708 & 0.521 & 126.825 & 199.676 & 146.937 \\
          &     & 49.3 & $W^{H}$ & 0.820 & 0.883 & 124.406 & 213.754 & 300.193 \\
          &     &      & $V^{F}$ & 0.743 & 0.574 & 123.302 & 193.682 & 149.628 \\
\noalign{\smallskip}\hline \hline
\end{tabular}
\end{center}
\end{table*}

\section{Breakup reactions of $^{11}$Be}
\label{s:break}

\subsection{The $^{10}$Be+$n$ model of $^{11}$Be}

In this section we consider the characteristics of breakup
processes of the $^{11}$Be nucleus, namely diffraction and
stripping reaction cross sections and the momentum  distributions
of the fragments. We use a simple model in which $^{11}$Be
consists of a core of $^{10}$Be and a halo of a single neutron
(see, e.g., \cite{Fukuda2004}). In this model the density of
$^{10}$Be has to be given. As in Sect.~\ref{s:elastic} we use the
QMC \cite{Pieper2002} and GCM \cite{Descouvemont97} density
distributions of $^{10}$Be. The hybrid model is applied to
calculate the OP of the interaction of $^{10}$Be with the target,
as well as OP for the $n$+target interaction. In the final step of
the procedure the sum of these potentials is folded with the
respective density distribution corresponding to the relative
motion wave function of the clusters in $^{11}$Be. The latter is
obtained by solving the Schr\"{o}dinger equation with the
Woods-Saxon potential for a particle with a reduced mass of two
clusters. The parameters of the WS potentials are obtained by a
fitting procedure, namely, to reach the neutron separation energy
$S_n=504 \pm 6$ KeV. They have the following values for $2s$ state
in which the valence neutron in $^{11}$Be is mainly bound (see
Refs.~\cite{Hencken96,Esbensen96}): $R=2.7$ fm, $a=0.52$ fm and
$V_0=61$ MeV. The rms radius of the cluster formation is obtained
to be 6.87 fm.

The $s$-state ($l=0, n=1,2$) of the relative motion of two
clusters has the form:
\begin{equation}\label{eq:6}
\phi_{00}^{(n)}({\bf s})=\phi_{0}^{(n)}(s)\frac{1}{\sqrt{4\pi}},
\;\;\; n=1,2 \; .
\end{equation}
The corresponding density distribution is the probability of both
clusters to be at a mutual distance $s$:
\begin{equation}\label{eq:7}
\rho_{0}^{(n)}({\bf s})=|\phi_{00}^{(n)}({\bf s})|^{2}= \frac{1}{4\pi}|\phi_{0}^{(n)}(s)|^{2}.
\end{equation}
Within  the $^{10}$Be+$n$ cluster model, in order to calculate the
$^{11}$Be breakup in its collision with the protons and nuclear
targets, one should calculate two OPs of $^{10}$Be+$p$(or $A$) and
$n$+$p$(or $A$) scattering:
\begin{widetext}
\begin{eqnarray}
U^{(b,n)}(r)&=&V^{(b,n)}+iW^{(b,n)}=\int d{\bf
s}\rho_{0}^{(n)}(s)\left [U_{c}^{(n)}\left ({\bf r}+ (1/11){\bf
s}\right )+U_{n}^{(n)}\left ({\bf r}-(10/11){\bf
s}\right )\right ]=2\pi \int_{0}^{\infty} \rho_{0}^{(n)}(s)s^{2}ds \nonumber \\
&\times & \int_{-1}^{1} dx \left
[U_{c}^{(n)}\left(\sqrt{r^{2}+(1s/11)^{2}+r(2/11)sx}\right )+
U_{n}^{(n)}\left(\sqrt{r^{2}+(10s/11)^{2}-r(20/11)sx}\right )\right
].
\label{eq:8}
\end{eqnarray}
\end{widetext}
In Eq.~(\ref{eq:8}) ${\bf r}-(10/11){\bf s}\equiv {\bf r}_{n}$ and
${\bf r}+(1/11){\bf s}\equiv {\bf r}_{c}$ give the distances
between the centers of each of the clusters and the target, and
${\bf s}={\bf s}_{1}+{\bf s}_{2}=(10/11){\bf s}+(1/11){\bf s}$
determines the relative distance between the centers of the two
clusters. $s_{1}$ and $s_{2}$ are the distances between the
centers of $^{11}$Be and each of the clusters, correspondingly.
The respective OPs for the $^{10}$Be+$A$ and $n$+$A$ scattering
are calculated within the microscopic model of OP from Sect.~II.A.

In the case of the $^{11}$Be breakup on the proton target the
$n$+$p$ potential is taken in the form \cite{Thompson77} (in MeV):
\begin{equation}
U_{n}^{(n)}=v_{np}=v(r)(1+i\gamma).
\label{eq:9}
\end{equation}
with
\begin{equation}
v(r)=120e^{-1.487r^{2}}-53.4e^{-0.639r^{2}}-27.55e^{-0.465r^{2}},
\label{eq:10}
\end{equation}
where $\gamma=0.4$.

For calculations of breakup cross sections and momentum
distributions of fragments in the $^{10}$Be+$n$ breakup model we
give here briefly the eikonal formalism, namely the expressions of
the $S$-matrix (as a function  of the impact parameter $b$):
\begin{equation}
S(b)=\exp \left [-\frac{i}{\hbar v}\int_{-\infty}^{\infty}
U(\sqrt{b^{2}+z^{2}}) dz \right ], \label{eq:11}
\end{equation}
where
\begin{equation}
U=V+ i W
\label{eq:12}
\end{equation}
is the OP. For negative $V$ and $W$ one can write
\begin{eqnarray}
S(b)&=& \left[ \cos \left( \frac 1 {\hbar v}\int_{-\infty}^\infty
|V| dz \right) + i \sin \left( \frac 1 {\hbar
v}\int_{-\infty}^\infty |V| dz \right) \right] \nonumber \\
& \times & \exp \left [ -\frac{1}{\hbar v}\int_{-\infty}^\infty
|W|dz \right ] , \label{eq:13}
\end{eqnarray}
and, correspondingly,
\begin{equation}
|S(b)|=\exp \left [ -\frac{1}{\hbar v}\int_{-\infty}^\infty |W| dz
\right ]. \label{eq:14}
\end{equation}
In our case $W$ is the imaginary part of the microscopic OP
[Eq.~(\ref{eq:8})]. $|S(b)|^2$ gives the probability that after
the collision with a proton ($z\rightarrow \infty$) (in the
$^{11}$Be+$p$ scattering), the cluster $c$ or the neutron with
impact parameter $b$ remains in the elastic channel ($i=c,n$):
\begin{equation}
|S_{i}(b)|^{2}=\exp{\left[-\frac{2}{\hbar
v}\int_{-\infty}^{\infty} dz\,\left |W_i
(\sqrt{b^{2}+z^{2}})\right |\right ]}.
\label{eq:15}
\end{equation}
The probability a cluster to be removed from the elastic channel
is $(1-|S|^{2})$. The probability of the case when both clusters
($c$ and $n$) leave the elastic channel
is $(1-|S_{n}|^{2})(1-|S_{c}|^{2})$. As shown in the next
subsection, Eqs.~(\ref{eq:11})-(\ref{eq:15}) take part in the
calculations of the diffraction breakup and stripping reaction
cross sections.

\subsection{Momentum distributions of fragments}

The necessary quantity to calculate the diffraction breakup and
absorption scattering cross sections (differential and total) and
momentum distributions is the probability function of the ${\bf
k}$-momentum distribution of a cluster in the system of two
clusters as a function of the impact parameter $\bf b$
\cite{Hencken96}:
\begin{equation}
\frac{d^{3}P_{\Omega}({\bf b},{\bf k})}{d{\bf k}}=\frac{1}{(2\pi)^{3}}
\left |\int d{\bf s} \phi_{{\bf k}}^{*}({\bf s})\Omega({\bf b},{\bf r}_{\perp})
\phi_{00}^{(n)}({\bf s})\right |^{2}.
\label{eq:21}
\end{equation}
In Eq.~(\ref{eq:21}) $\Omega({\bf b},{\bf r}_{\perp})$ is given by
the products of two $S$-functions $S_{c}$ and $S_{n}$
[Eqs.~(\ref{eq:11})-(\ref{eq:15})] of the core $^{10}$Be and the
neutron, $\phi_{{\bf k}}({\bf s})$ is the continuum wave function,
${\bf k}$ is the relative momentum of both clusters in their
center-of-mass frame, and the vector ${\bf r}_{\perp}$ is the
projection of the relative coordinate ${\bf s}$ between the
centers of the two clusters on the plane normal to the $z$-axis.
The bound-state wave function $\phi_{00}$ of the relative motion
of two clusters  is given for the $s$-state by Eq.~(\ref{eq:6}).
As to the wave function in the final state $\phi_{{\bf k}}$, we
will neglect its distortion and, thus, replace it by $j_{0}(ks)$
in the case of the $s$-state. Then, following
Ref.~\cite{Hencken96}, the probability function has the form
\begin{widetext}
\begin{equation}
\frac{d^{2}P_{\Omega}({\bf b},{\bf k})}{dk_{L}dk_{\perp}}=
\frac{k_{\perp}}{16\pi^{3}k^{2}}\left |\int ds \int
d(\cos\theta_{s})\,g(s)\sin{(ks)}\int d\varphi_{s}\Omega({\bf
b},{\bf r}_{\perp})\right |^{2} ,
\label{eq:22}
\end{equation}
\end{widetext}
where
\begin{equation}
\Omega({\bf b},{\bf r}_{\perp})=S_{c}({\bf b}_{c})S_{n}({\bf
b}_{n})
\label{eq:23}
\end{equation}
and $g(s)=s\phi_{0}^{(n)}(s)$, $\phi_{0}^{(n)}$ being given by
Eq.~(\ref{eq:6}).

Hence, the diffraction breakup cross section has the form
%
\begin{equation}
\left (\frac{d\sigma}{dk_{L}}\right
)_{\text{diff}}=\int_{0}^{\infty} b_n db_{n}\int_{0}^{2\pi} d
\varphi_{n}\int_{0}^{\infty} d{k}_{\perp}
\frac{d^{2}P_{\Omega}({\bf k},{\bf b})}{dk_{L} dk_{\perp}} .
\label{eq:24}
\end{equation}
%
In Eq.~(\ref{eq:24}) $d^{2}P_{\Omega}({\bf b},{\bf
k})/dk_{L}dk_{\perp}$ is given by Eq.~(\ref{eq:22}). The integrals
over $b_n$ and $\varphi_{n}$ mean integration over the impact
parameter ${\bf b}_n$ of the neutron with respect to the target.

The cross sections of the stripping reaction when the neutron
leaves the elastic channel is \cite{Hencken96}:
\begin{eqnarray}
\left(\frac{d\sigma}{dk_{L}}\right)_{\text{str}}&=&\frac{1}{2\pi^{2}}\int_{0}^{\infty}b_{n}d\,b_{n}d\varphi_{n}
\left [ 1-|S_{n}(b_{n})|^{2}\right ] \nonumber \\
& \times & \int \rho d\rho d\varphi_{\rho} |S_{c}(b_{c})|^{2}
\nonumber \\
& \times & \left [ \int_{0}^{\infty}dz \cos (k_{L}z)\phi_{0}\left
(\sqrt{\rho^{2}+z^{2}}\right )\right]^{2}.
\label{eq:25}
\end{eqnarray}
Equation (\ref{eq:25}) is obtained when the incident nucleus has
spin equal to zero and for the $s$-state of the relative motion of
two clusters in the nucleus with ${\bf s}={\bf r}_{c}- {\bf
r}_{n}$, ${\bf \rho}={\bf b}_{c}-{\bf b}_{n}$, ${\bf s}={\bf \rho}
+ {\bf z}$ and
\begin{equation}
b_{c}=\sqrt{s^{2}\sin^{2}\theta+b_{n}^{2}+2sb_{n}\sin\theta\cos(\varphi-
\varphi_{n})}
\label{eq:18}
\end{equation}
coming from ${\bf b}_{c}={\bf b}_{n}+{\bf b}$, where
$b=s\sin\theta$ is the projection of ${\bf s}$ on the plane normal
to the $z$-axis along the straight-line trajectory of the incident
nucleus.

In the end of this subsection we note that the real and imaginary
parts of the OPs taking part in Eq.~(\ref{eq:8}) and in the
$S$-matrices [Eqs.~(\ref{eq:11})-(\ref{eq:15})] are used for
calculations of the cross sections
[Eqs.~(\ref{eq:21})-(\ref{eq:18})] in the cases of scattering and
breakup of $^{11}$Be on protons and nuclei that will be considered
in the following part of our work. They are calculated
microscopically within the hybrid model given in subsection II.A.

\subsection{Results of calculations of breakup reactions}

\begin{figure}
\includegraphics[width=0.9\linewidth]{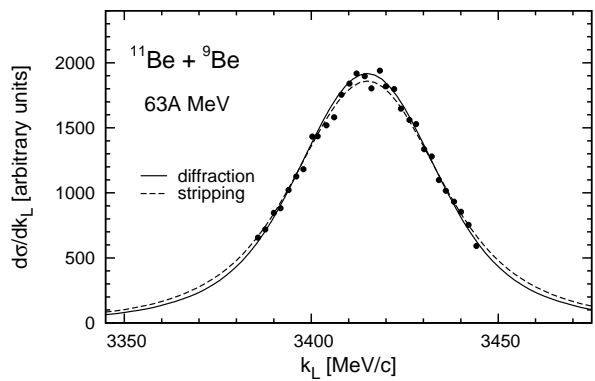}
\caption{Cross sections of diffraction breakup and stripping
reaction in $^{11}$Be+$^{9}$Be scattering at $E=63$ MeV/nucleon.
Experimental data are taken from Ref.~\protect\cite{Kelley95}.
\label{fig7}}
\end{figure}

\begin{figure}
\includegraphics[width=0.9\linewidth]{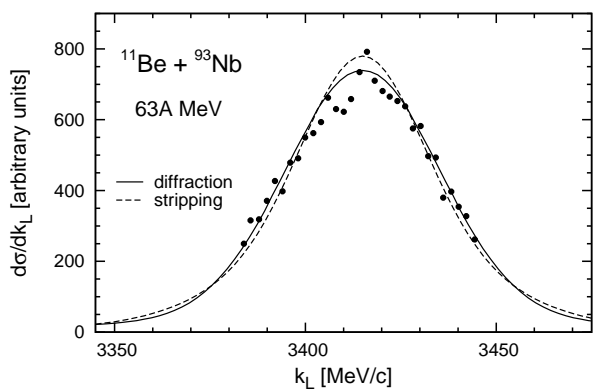}
\caption{The same as Fig.~\ref{fig7}, but for $^{11}$Be+$^{93}$Nb
scattering.
\label{fig8}}
\end{figure}

\begin{figure}
\includegraphics[width=0.9\linewidth]{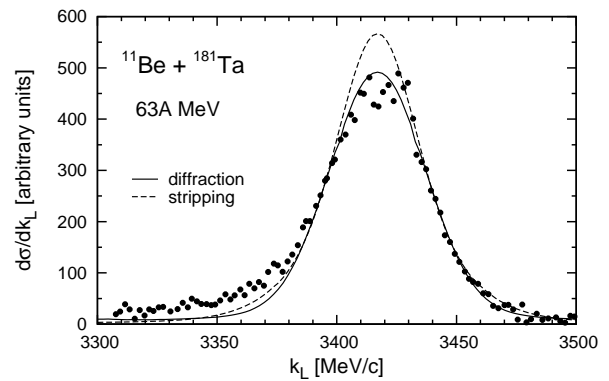}
\caption{The same as Fig.~\ref{fig7}, but for $^{11}$Be+$^{181}$Ta
scattering.
\label{fig9}}
\end{figure}

\begin{figure}
\includegraphics[width=0.9\linewidth]{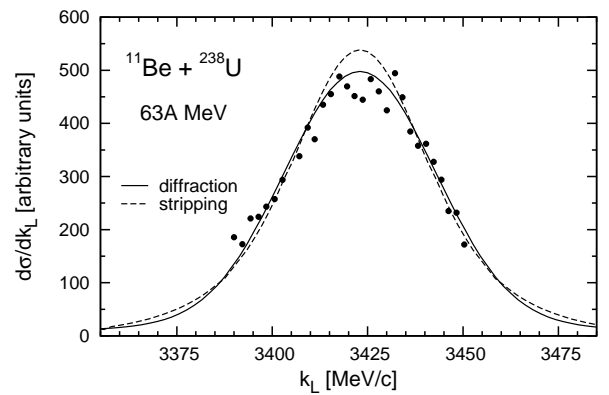}
\caption{The same as Fig.~\ref{fig7}, but for $^{11}$Be+$^{238}$U
scattering.
\label{fig10}}
\end{figure}

In this subsection we perform calculations of the breakup cross
sections of $^{11}$Be on the target nucleus $^{9}$Be and heavy
nuclei, such as $^{93}$Nb, $^{181}$Ta, and $^{238}$U, and compare
our results with the available experimental data \cite{Kelley95}.
The densities of these heavy nuclei needed to compute the OPs are
taken from Ref.~\cite{Patterson2003}. The diffraction and
stripping cross sections (when a neutron leaves the elastic
channel) for reactions $^{11}$Be+$^{9}$Be, $^{11}$Be+$^{93}$Nb,
$^{11}$Be+$^{181}$Ta, and $^{11}$Be+$^{238}$U are calculated from
Eqs.~(\ref{eq:24}) and (\ref{eq:25}). The obtained results are
illustrated in Figs.~\ref{fig7}, \ref{fig8}, \ref{fig9}, and
\ref{fig10}, respectively. We note the good agreement with the
experimental data from light and heavy breakup targets. The
obtained cross sections for the diffraction and stripping have a
similar shape. The values of the widths are around $50$ MeV in
agreement with the experimental ones. Our results confirm the
observations (e.g., in Refs.~\cite{Orr92,Orr95}) that the width
almost does not depend on the mass of the target and as a result,
it gives information basically about the momentum distributions of
two clusters. Here we note that due to the arbitrary units of the
measured cross sections of the considered processes it was not
necessary to renormalize the depths of our OPs of the
fragments-target nuclei interactions.


\section{Conclusions}
\label{s:summary}

In the present work the hybrid model is applied to study
characteristics of the processes of scattering and reactions of
$^{10}$Be and $^{11}$Be on protons and nuclei. In the model, the
ReOP is calculated  microscopically in a folding procedure of the
densities of the projectile and the target with  effective NN
interactions related to the $g$-matrix obtained on the basis of
the Paris NN potential. The ReOP includes both the direct and
exchange terms. The ImOP is calculated microscopically as the
folding OP that reproduces the phase of scattering obtained in the
high-energy approximation. The only free parameters in the hybrid
model ($N$s) are the coefficients that correct the depths of the
ReOP, ImOP, and the spin-orbit parts of OP. Their values are
obtained by a fitting procedure to the experimental data whenever
they exist. Additionally, in some cases the surface absorption is
accounted for by including another term to the OP that requires
one more fitting parameter. The density distributions of $^{10}$Be
obtained within GCM and QMC microscopic methods and of $^{11}$Be
from GCM are used. The resulting within the hybrid model OPs are
applied to calculate characteristics of various processes.

The results of the present work can be summarized as follows.

(i) Elastic scattering cross sections of $^{10}$Be and $^{11}$Be
on protons and $^{12}$C are calculated using the microscopic OPs
for energies $E<100$ MeV/nucleon and compared with the existing
experimental data. In order to resolve the ambiguities of the
magnitudes of the depths of the OPs, the well established energy
dependence of the respective volume integrals of the OPs is taken
into account. The theoretical approach gives a good explanation of
a wide range of empirical data on the $^{10,11}$Be+$p$ and
$^{10,11}$Be+$^{12}$C elastic scattering. It was established that
the obtained by fitting procedure values of the coefficients $N_R$
(depths of ReOP) are close to unity. The correction of the ImOP by
factor $N_I$ is in some cases larger, e.g., for $^{10}$Be+$p$ at
energy $E=39.1$ MeV/nucleon in the case when the spin-orbit ($ls$)
component is not accounted for. The inclusion of a surface term to
the OP leads to a better agreement with the experimental elastic
scattering cross section data. We conclude that, in general, the
hybrid model for microscopic calculations of the OPs gives the
basic important features of the scattering cross sections and can
be recommended and applied to calculate more complex processes
such as breakup reactions, momentum distributions of fragments and
others.

(ii) Apart from the usual folding model, we use another folding
approach to consider the $^{11}$Be breakup by means of the simple
$^{10}$Be+$n$ cluster model for the structure of $^{11}$Be. Within
this folding model we construct the OP of the interaction of
$^{10}$Be with the target, as well as the $n$+target interaction.
Using the cluster OPs $^{10}$Be+$p$(or $A$) and $n$+$p$(or $A$)
the corresponding functions $S_c$ and $S_n$ ($S$-matrices) for the
core and neutron within the eikonal formalism are obtained.

(iii) The calculated $S_c$ and $S_n$ functions are used to get
results for the longitudinal momentum distributions of $^{10}$Be
fragments produced in the breakup of $^{11}$Be on different
targets. This includes the breakup reactions of $^{11}$Be  on
$^{9}$Be, $^{93}$Nb, $^{181}$Ta and $^{238}$U at $E=63$
MeV/nucleon for which a good agreement of our calculations for the
diffraction and stripping reaction cross sections with the
available experimental data exist. The obtained widths of about  0
MeV/c are close to the empirical ones. Future measurements of such
reactions are highly desirable for the studies of the exotic
$^{11}$Be structure. The accurate interpretation of the expected
data requires more refined theoretical approaches, for instance
that of Ref.~\cite{Diego2014} within the CDCC method and its
extensions to study the effects of the dynamic core excitation,
especially its large contribution to nuclear breakup in the
scattering of halo nuclei.

\begin{acknowledgments}
The authors are grateful to S. C. Pieper for providing with the
density distributions of $^{9,10}$Be nuclei calculated within the
QMC method. The work is partly supported by the Project from the
Agreement for co-operation between the INRNE-BAS (Sofia) and JINR
(Dubna). Four of the authors (D.N.K., A.N.A., M.K.G. and K.S.) are
grateful for the support of the Bulgarian Science Fund under
Contract No.~DFNI--T02/19 and one of them (D.N.K.) under Contract
No.~DFNI--E02/6. The authors V.K.L., E.V.Z., and K.V.L. thank the
Russian Foundation for Basic Research (Grant No. 13-01-00060) for
partial support. K.S. acknowledges the support of the Project
BG-051P0001-3306-003.
\end{acknowledgments}

\end{document}